# Intrinsic Instability of the Hybrid Halide Perovskite Semiconductor $CH_3NH_3PbI_3$

Yue-Yu Zhang[1], Shiyou Chen[2*], Peng Xu[1], Hongjun Xiang[1], Xin-Gao Gong[1*], Aron Walsh[3], and Su-Huai Wei[4]

[1]Key Laboratory for Computational Physical Sciences (MOE), State Key Laboratory of Surface Physics, and Department of Physics, Fudan University, Shanghai 200433, China

[2]Key Laboratory of Polar Materials and Devices (MOE), East China Normal University, Shanghai 200241, China

[3]Center for Sustainable Chemical Technologies and Department of Chemistry, University of Bath, Bath BA2 7AY, UK

[4]National Renewable Energy Laboratory, Golden, Colorado 80401, USA

**Abstract**

The organic–inorganic hybrid perovskite $CH_3NH_3PbI_3$ has attracted significant interest for its high performance in converting solar light into electrical power with an efficiency exceeding 20%. Unfortunately, chemical stability is one major challenge in the development of the $CH_3NH_3PbI_3$ solar cells. It was commonly assumed that moisture or oxygen in the environment causes the poor stability of hybrid halide perovskites, however, here we show from the first-principles calculations that the room-temperature tetragonal phase of $CH_3NH_3PbI_3$ is thermodynamically unstable with respect to the phase separation into $CH_3NH_3I + PbI_2$, i.e., the disproportionation is exothermic, independent of the humidity or oxygen in the atmosphere. When the structure is distorted to the low-temperature orthorhombic phase, the energetic cost of separation increases, but remains small. Contributions from vibrational and configurational entropy at room temperature have been considered, but the instability of $CH_3NH_3PbI_3$ is unchanged. When I is replaced by Br or Cl, Pb by Sn, or the organic cation $CH_3NH_3$ by inorganic Cs, the perovskites become more stable and do not phase-separate spontaneously. Our study highlights that the poor chemical stability is intrinsic to $CH_3NH_3PbI_3$ and suggests that element-substitution may solve the chemical stability problem in hybrid halide perovskite solar cells.

**-Introduction**

Inorganic-organic hybrid perovskite compounds ($CH_3NH_3PbX_3$, X=I, Br and Cl) have been intensively studied as light-harvesting semiconductors in solar cells because of their strong optical absorption and high carrier mobility[1-9]. The power conversion efficiency (PCE) increases rapidly in the past three years, and now it is over 20%[10,11], close to the record efficiency of the conventional silicon crystal[12,13], CdTe[14] and $Cu(In,Ga)Se_2$[15] thin film solar cells which have been studied for several decades.

Despite the competitive photovoltaic efficiency, a major challenge is the poor material stability, which remains an obstacle in the development of commercially viable $CH_3NH_3PbI_3$ solar cells[16-22]. The degradation process of perovskite-structured $CH_3NH_3PbI_3$ can occur easily in humid environments, and thus the device fabrication should be carried out with a humidity <1%, as suggested by Grätzel and co-workers[2]. However, the microscopic origin and detailed process of the degradation is still not clear. The experiments of Niu *et al.* showed that moisture, oxygen and UV radiation play a role in the degradation progress of perovskite $CH_3NH_3PbI_3$[23]. On the other hand, Schoonman proposed that the Pb-I components of the perovskite structure may exhibit photodecomposition similar to that of the binary halides[24]. This process is easy to understand, as the upper valence band is formed by the antibonding states of the Pb 6s-I 5p hybridization[25,26], which increases the dispersion of the valence bands (thus small hole effective masses[27]) but also weakens the Pb-I bonds.

Although the poor stability of $CH_3NH_3PbI_3$ can be understood from different perspectives, it is generally assumed that $CH_3NH_3PbI_3$ is a stable compound with respect to the phase separation, i.e., it will not disproportionate spontaneously[28]. If this assumption is true, the degradation of the $CH_3NH_3PbI_3$ solar cells can be suppressed if the compound is protected from the moisture, oxygen and light-illumination. In contrast, we demonstrate that $CH_3NH_3PbI_3$ in the room-temperature tetragonal structure is thermodynamically unstable and phase-separation into the $CH_3NH_3I + PbI_2$ is an exothermic process. Therefore, the long-term stability is questionable even if the samples are protected from the environment. The thermodynamic stability increases in the low-temperature orthorhombic perovskite structure, but the energy cost for the phase-separation remains small. In contrast, the energy cost of separation increases if the Pb cation is replaced by Sn or $CH_3NH_3$ by Cs, which demonstrates that the thermodynamic potential may be tuned chemically for enhanced stability..

**-Results**

**Instability of $CH_3NH_3PbI_3$**

Similar to many $ABO_3$ perovskite oxides, $CH_3NH_3PbI_3$ has three temperature-dependent phases: high-temperature cubic (above 327.4 K)[29], room-temperature tetragonal (162.2-327 K)[30], and low-temperature orthorhombic structures[29], as shown in Fig. 1. The main differences between these structures are the distortion of the Pb-I sublattice and the disorder in the $CH_3NH_3^+$ sublattice. Here we consider the stability of the room-temperature tetragonal structures first. Previous experiments reported two tetragonal structures of $CH_3NH_3PbI_3$ with different space group, one in I4/mcm[29,30] and the other one in I4cm[30]. The total energy of the I4/mcm phase is calculated to be 20 meV/f.u. (by the vdW-TS method) lower than the I4cm phase, and the small energy difference indicates that they may coexist. In the following calculation, the I4/mcm phase with lower energy is considered as the tetragonal phase. Although the structure of Pb-I sublattice is determined experimentally in the two tetragonal structures, there is still another structural degree of freedom in $CH_3NH_3PbI_3$, the orientation of the polar $CH_3NH_3^+$ molecular cation. This is different from the inorganic perovskites such as $CsPbI_3$ in which the $Cs^+$ cation has no orientation freedom. We constructed a series of structures with different $CH_3NH_3^+$ orientations and identified the lowest-energy configuration, which is close to that determined through simulated annealing method by Agiorgousis et al[31]. The orientation is also compared to the orientation reported in other theoretical studies, showing that the orientation change may influence the energy by over 40 meV/f.u. The orientation with the lowest energy in our simulation cell is considered in the following discussion.

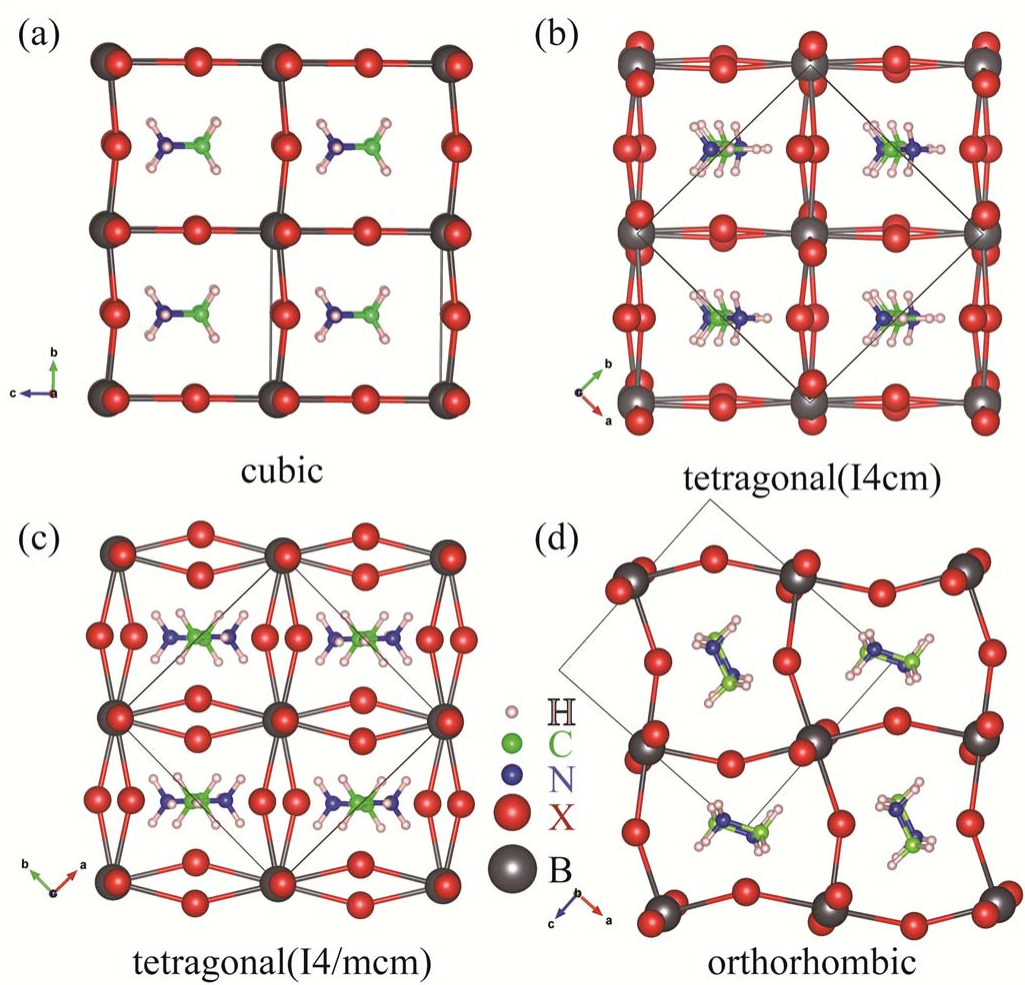

Figure 1 Representation of the crystal structures of (a) cubic, (b) tetragonal (space group: I4cm), (c) tetragonal (I4/mcm) and (d) orthorhombic $CH_3NH_3BX_3$ (B=Pb, Sn; X=Cl, Br, I) perovskites. The B, X, N, C, and H atoms are presented by black, red, blue, green and pink spheres, in order.

With the lowest-energy orientation determined for the tetragonal $CH_3NH_3PbI_3$ structure, we now study its stability with respect to the phase separation,

$$CH_3NH_3PbI_3 \rightarrow CH_3NH_3I + PbI_2 \quad (1)$$

Unexpectedly, the calculated energy change (listed in Table 1) shows that this reaction is exothermic (thermodynamically favorable), so it may occur spontaneously even without any moisture, oxygen or illumination in the environment, which imposes a serious limit on the stabilization of the $CH_3NH_3PbI_3$ solar cells. From the thermodynamic point of view, this phase-separation reaction cannot be avoided, so the long-term stability is always poor. It should be noted that the kinetic barrier may prevent the compound from phase-separation once it is formed, so $CH_3NH_3PbI_3$ may still be stable for a certain period, which explains the fact that the synthesized $CH_3NH_3PbI_3$ samples are stable and can work as an efficient solar cell light-harvesting material for several days[10]. As a result of the intrinsic instability, increasing the kinetic barrier of the phase-separation becomes crucial to suppressing the degradation of $CH_3NH_3PbI_3$ solar cells.

Table 1. The calculated energy cost (in eV/f.u.) of the phase-separation reactions of the hybrid halide perovskites in three crystal polymorphs and with two exchange-correlation functionals (PBE and vdW-TS).

| Phase-Separation | cubic | | tetragonal | | orthorhombic | |
|---|---|---|---|---|---|---|
| | PBE | vdW | PBE | vdW | PBE | vdW |
| $CH_3NH_3PbI_3 \rightarrow CH_3NH_3I + PbI_2$ | -0.111 | -0.119 | -0.060 | -0.063 | -0.031 | 0.037 |
| $CH_3NH_3PbBr_3 \rightarrow CH_3NH_3Br + PbBr_2$ | 0.043 | 0.014 | 0.077 | 0.065 | 0.068 | 0.106 |
| $CH_3NH_3PbCl_3 \rightarrow CH_3NH_3Cl + PbCl_2$ | 0.040 | 0.004 | 0.058 | 0.033 | 0.097 | 0.071 |
| $CH_3NH_3SnI_3 \rightarrow CH_3NH_3I + SnI_2$ | 0.070 | 0.076 | 0.248 | 0.129 | 0.239 | 0.141 |
| $CH_3NH_3SnBr_3 \rightarrow CH_3NH_3Br + SnBr_2$ | 0.281 | 0.140 | 0.281 | 0.148 | 0.286 | 0.176 |
| $CH_3NH_3SnCl_3 \rightarrow CH_3NH_3Cl + SnCl_2$ | 0.287 | 0.126 | 0.288 | 0.136 | 0.299 | 0.174 |
| $CsPbI_3 \rightarrow CsI + PbI_2$ | -0.069 | | | | 0.098 | |
| $CsPbBr_3 \rightarrow CsBr + PbBr_2$ | 0.127 | | | | 0.209 | |
| $CsPbCl_3 \rightarrow CsCl + PbCl_2$ | 0.224 | | | | 0.292 | |
| $CsSnI_3 \rightarrow CsI + SnI_2$ | 0.115 | | | | 0.201 | |
| $CsSnBr_3 \rightarrow CsBr + SnBr_2$ | 0.259 | | | | 0.293 | |
| $CsSnCl_3 \rightarrow CsCl + SnCl_2$ | 0.324 | | | | 0.335 | |

Similar to the hybrid perovskites, there is orientational freedom in the structure of $CH_3NH_3I$[15,16], so we considered the energy dependence on the $CH_3NH_3$ orientation (the change is only 4 meV/f.u., smaller than that of $CH_3NH_3PbI_3$) and used the lowest-energy orientation when calculating the energy cost of Reaction (1). Furthermore, the distance between organic ($CH_3NH_3$) and inorganic ($PbI_3$) components in the perovskite $CH_3NH_3PbI_3$ is large and they are bound partially by the van der Waals interaction, so we used both PBE and vdW-TS approximated exchange-correlation functionals to relax the crystal structures. In general, the PBE functional overestimates the lattice constants while the vdW-TS results agree better with the experiment results, which is consistent with the calculation by Wang et al. using optb86B vdW functional[32]. The improvement can be attributed to that the vdW-TS functional describes the dispersion interaction between the organic components and inorganic framework more accurately. Using both functionals, the calculated energy change of Reaction (1) is always negative. Furthermore, other functionals including the local density approximation and PBEsol have also been used and the calculated energy cost is also negative. This indicates that the conclusion is not influenced by the specific approximations to the exchange-correlation functional.

Although the room-temperature tetragonal phase of $CH_3NH_3PbI_3$ is not stable with respect to phase separation, it is still a question whether the low-temperature orthorhombic phases is stable. As expected, the calculated energies of the low-symmetry orthorhombic and tetragonal structures and high-symmetry cubic structures increase in order, which is common in many perovskite oxides. At low temperature, the lowest-energy orthorhombic structure is dominant, while at higher temperature, the higher-energy tetragonal and cubic structures appear[29,33,34]. Since the cubic structure has higher energy than the tetragonal one, the energy cost of its phase separation is more negative (as shown in Table 1), so the tendency for phase separation is even stronger. The orthorhombic structure has lower energy than the tetragonal one, so its phase separation costs more energy. The calculated energy cost depends on the specific quantum mechanical treatment, i.e., the PBE result is weakly exothermic, so the orthorhombic structure is still not stable, while the vdW-TS result is weakly positive, so it is stable and will not phase-separate. Because the calculated energy cost is always small for the orthorhombic structure, the phase separation may still occur on the surface, where the crystal energy is higher and kinetic barriers are lowered. It has recently been shown that $H_2O$ can effectively intercalate the lattice[35], which increases the internal surface area. This is a possible microscopic mechanism for the observation that the water (moisture) can catalyze the phase separation and degradation of the $CH_3NH_3PbI_3$ solar cells.

**Stability Enhancement by Element Substitution**

To improve the stability of $CH_3NH_3PbI_3$ solar cells, one possible method is to substitute the organic $CH_3NH_3^+$ and inorganic $Pb^{2+}$ cations, or $I^-$ anions by similar elements, e.g., replacing $Pb^{2+}$ by $Sn^{2+}$, $I^-$ by $Br^-$, $Cl^-$, or $CH_3NH_3^+$ by $Cs^+$ and forming a series of $ABX_3$ (A= $CH_3NH_3$, Cs, B=Pb, Sn, X=I, Br, Cl) compounds. These elements may have different binding with each other, so it is hopeful that the stability of the $ABX_3$ may be enhanced relative to the phase-separated AX + $PbX_2$ compounds. The calculated energetic cost for phase separation of these compounds is also listed in Table 1. When $I^-$ is replaced by $Br^-$, the reactions become endothermic for all three structures (with both PBE and vdW calculations). Therefore $CH_3NH_3PbBr_3$ will not phase-separate at low temperature. However, the energy cost is still small, so the long-term stability of $CH_3NH_3PbBr_3$ is still limited. The situation in $CH_3NH_3PbCl_3$ is at variance with that in $CH_3NH_3PbBr_3$. The calculated energy cost is more positive than that of $CH_3NH_3PbBr_3$ from the PBE calculation, however, it is less positive from the vdW calculation. Here, the stronger ionicity of $CH_3NH_3PbCl_3$ may not be described accurately in the vdW-TS calculations. Comparing the PBE results of $CH_3NH_3PbBr_3$ and $CH_3NH_3PbCl_3$, we can find that they are similar and small. It can be predicted that although the mixed phase systems, $CH_3NH_3Pb(I_{1-x}Cl_x)_3$ and $CH_3NH_3Pb(I_{1-x}Br_x)_3$, are expected to be more stable than $CH_3NH_3PbI_3$,[5,36] their long-term stability is still poor.

When Pb is replaced by non-toxic Sn, enhancement of the stability becomes possible with respect to Sn(II) salts. The calculated energy cost of the phase-separation reaction increases to more than 0.2 eV (PBE result), much higher than those (all less than 0.1 eV or even negative) of the Pb compounds. The larger energy cost indicates that the $CH_3NH_3SnI_3$, $CH_3NH_3SnBr_3$ and $CH_3NH_3SnCl_3$ will have a lesser tendency for phase separation. We noticed that $CH_3NH_3SnI_3$ was found to have poor stability because of the instability of Sn(II)[37,38] itself with respect to oxidation. Our study confirms that the poor stability of $CH_3NH_3SnI_3$ does not result from disproportionation (the valence of Sn is not changed in the reaction discussed here), but is instead associated with Sn(II) to Sn(IV) oxidation processes.

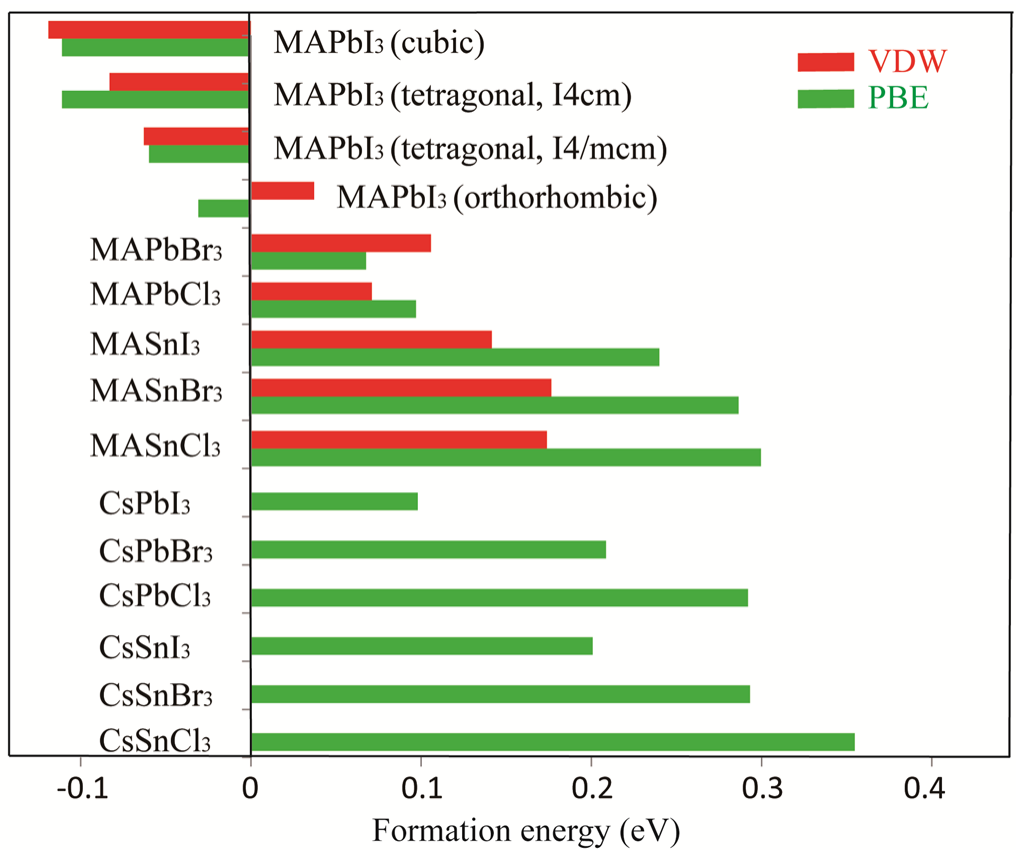

Figure 2 The calculated energy cost of the phase separation of $ABX_3$ (A= $CH_3NH_3$, Cs, B=Pb, Sn, X=I, Br, Cl) in their orthorhombic structure. For $CH_3NH_3PbI_3$, the results of all the three structures are plotted. For the organic-inorganic hybrid perovskites, the results from both the vdW-TS and PBE functionals are plotted. Positive number indicates the compound is stable at T = 0 K.

Enhancement of the stability can be also achieved when the organic cation $CH_3NH_3^+$ is replaced by the inorganic $Cs^+$. As shown in Table 1 and Figure 2, the energy cost of $CsPbI_3$ phase separation is as high as 98 meV/f.u., which means that the fully inorganic compound will not phase-separate under low or even room temperature, and a much better long-term stability may be achieved. Comparing $CH_3NH_3PbI_3$ and $CsPbI_3$, the former one has larger band gap because of the larger size of the organic cation and also has benign defect properties (no deep-level defects)[31,39-41], so it has high solar cell efficiency. In contrast, the inorganic perovskite has lower efficiency because of its smaller gap and deep-level defects[42], but its stability is better. This tradeoff should be considered if one intends to enhance the stability of $CH_3NH_3PbI_3$ solar cells through mixing (alloying) the A cations.

**Stability at High Temperature**

So far the stability of the perovskites is predicted according to the calculated energy change (internal energy changes at zero pressure and zero temperature for each phase) following Reaction (1). The true thermodynamic stability is determined by the Gibbs free energy, which includes contributions from internal energy, pressure (which can be neglected for solids in air) and temperature (vibrational and configurational entropy). The entropy contribution can be significant at high temperatures. We will first discuss the influence of the vibrational entropy.

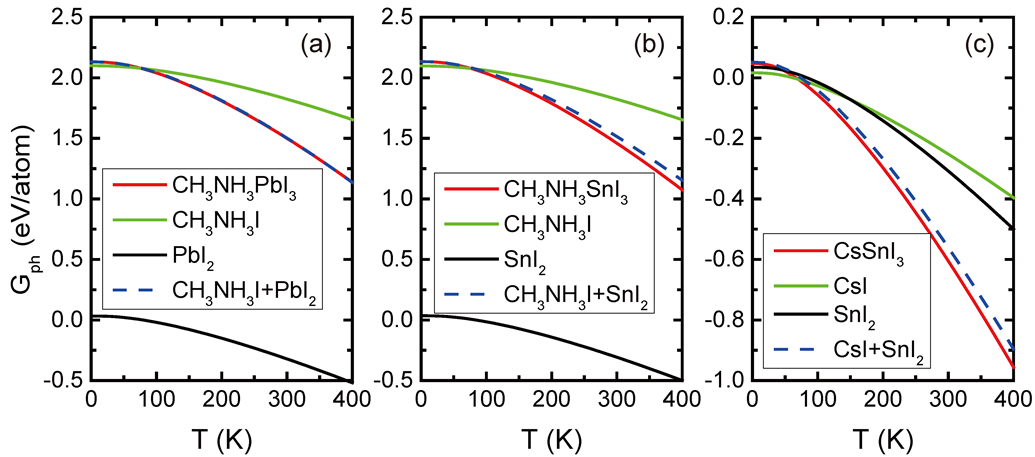

Figure 3 The calculated vibrational free energy of (a) $CH_3NH_3PbI_3$, (b) $CH_3NH_3SnI_3$ and (c) $CsSnI_3$ and their competitive compounds.

The vibration contribution to the Gibbs free energy $G_{ph}(T)$ can be calculated based on the phonon spectrum (harmonic approximation)[43]

$$G_{ph}(T) = \sum_{q,v} \frac{\hbar \omega_{q,v}}{2} + k_B T \sum_{q,v} \ln\left[1 - \exp\left(-\frac{\hbar \omega_{q,v}}{k_B T}\right)\right] \quad (2)$$

where $T$ is the temperature, $q$ and $v$ are the wave vector and band index, respectively, and $\omega_{q,v}$ is the phonon frequency at $q$ and $v$. $k_B$ and $\hbar$ are the Boltzmann constant and the reduced Planck constant. The first term is the zero-point vibration energy, and the second term is the vibrational free energy. In Fig. 3(a), $G_{ph}$ is plotted as a function of the temperature for $CH_3NH_3PbI_3$ and the phase-separated $CH_3NH_3I$ + $PbI_2$. Interestingly, the vibrational contribution is comparable for the reactant and products of phase-separation, so it does not influence the energy cost or the stability of $CH_3NH_3PbI_3$. The negligible influence can be understood according to the character of the phonon spectrum, i.e., the high-frequency phonon modes come from the $CH_3NH_3$ molecule (the vibration of C-N, C-H, and N-H bonds), which is similar in the $CH_3NH_3PbI_3$ and $CH_3NH_3I$. Similar calculations have also been performed for $CH_3NH_3SnI_3$ and $CsSnI_3$. The vibration entropy enhances the stability of $CsSnI_3$ slightly, because the vibrational free energy of $CsSnI_3$ decreases faster than $CsI + SnI_2$ under high temperature.

Configurational entropy may also decrease the Gibbs free energy. For ordered crystal semiconductors, the configurational entropy is usually negligible. However, for the organic-inorganic hybrid perovskites the orientation freedom of $CH_3NH_3$ molecule increases the number of configurations and thus the entropy contribution. Because this molecular rotational freedom exists in both $CH_3NH_3PbI_3$ and $CH_3NH_3I$, the net effect is dampened. Our calculation shows that the energy difference of different orientations is only 4 meV in $CH_3NH_3I$, smaller than that in $CH_3NH_3PbI_3$, which indicates $CH_3NH_3I$ may have higher configurational entropy than $CH_3NH_3PbI_3$, further destabilizing the perovskite structure.

**-Conclusions**

The stability of hybrid halide perovskites $ABX_3$ (A= $CH_3NH_3$, Cs; B=Pb, Sn; X=I, Br, Cl) is studied with respect to the disproportionation into $AX + BX_2$ using first-principles calculations. The high-efficiency solar cell material $CH_3NH_3PbI_3$ in the tetragonal structure is found to be thermodynamically unstable and tends to phase-separate, regardless of humidity or oxygen in the atmosphere. The orthorhombic structure is slightly more stable, i.e., the phase-separation reaction can be endothermic, but the energy cost is very small and thus the thermodynamic stability remains poor. When Pb is replaced by Sn, or the organic cation $CH_3NH_3$ is replaced by inorganic Cs, the intrinsic stability of the perovskites phases is enhanced. Similar effects occur when the anion I is replaced by Br or Cl, but the stability enhancement is lesser. Our study shows that poor stability is inherent to the $CH_3NH_3PbI_3$ compound, rather than determined by the moisture, oxygen or illumination in the environment as previously assumed. We propose that appropriate element substitution may enhance the stability of the hybrid perovskite solar cells.

**-Computational Methods**

We employ density functional theory (DFT) for crystal structure optimisation and electronic structure calculations. The ion-electron interaction is treated by the projector augmented-wave (PAW) technique,[44] as implemented in the *Vienna ab initio simulation package* (VASP)[45]. Both the Perdew-Burke-Ernzerhof (PBE)[46] and the non-local vdW-TS[47] method with cutoff radius 30 Å, which describe the London dispersion interaction more accurately, are adopted. Particular care has been given to the calculation convergence criteria to ensure reliable energies. The energy cutoff of the plane-wave basis set is 500 eV. The 3D *k*-point mesh is generated by the Monkhorst-Pack scheme ($6 \times 6 \times 6$ for cubic phase of $ABX_3$ (A= Cs, $CH_3NH_3$, B=Pb and Sn, X= I, Br and Cl), $6 \times 6 \times 4$ for tetragonal phase; $8 \times 8 \times 6$ for tetragonal phase of $CH_3NH_3X$ with space group P4/nmm, $5 \times 5 \times 5$ for rocksalt phase). The lattice vectors and atomic positions are optimized according to the atomic forces, with a criterion that the calculated force on each atom is smaller than 0.01 eV/Å. Phonon calculations are performed by supercell approach. Real-space force constants of supercells are calculated in the density-functional perturbation theory (DFPT) as implemented in the VASP code,[45] and phonon frequencies are calculated from the force constants using the PHONOPY code[48]. Thermal properties are calculated from phonon frequencies on a sampling mesh in the reciprocal space ($8 \times 8 \times 8$ for both cases of $CH_3NH_3PbI_3$ and $CsSnI_3$).

-Acknowledgements

The work at Fudan University was supported by the Special Funds for Major State Basic Research, National Natural Science Foundation of China (NSFC) and International collaboration project. S.C. is supported by NSFC under grant No. 91233121, Shanghai Rising-Star Program (14QA1401500) and CC of ECNU. The work at Bath is supported by the Royal Society, the ERC and EPSRC (Grant No. EP/M009580/1 and EP/K016288/1). The work at NREL was funded by the U.S Department of Energy (DOE), under Contract No. DE-AC36-08GO28308.

-Contributions

S. C and X. G conceived the idea. Y. Z, S. C and P. X performed the calculations. All the authors analyzed the results and wrote the paper.

-Corresponding authors

Correspondence to: Shiyou Chen (chensy@ee.ecnu.edu.cn) and Xin-Gao Gong (xggong@fudan.edu.cn)